


\font\bigbold=cmbx12
\font\eightrm=cmr8
\font\sixrm=cmr6
\font\fiverm=cmr5
\font\eightbf=cmbx8
\font\sixbf=cmbx6
\font\fivebf=cmbx5
\font\eighti=cmmi8  \skewchar\eighti='177
\font\sixi=cmmi6    \skewchar\sixi='177
\font\fivei=cmmi5
\font\eightsy=cmsy8 \skewchar\eightsy='60
\font\sixsy=cmsy6   \skewchar\sixsy='60
\font\fivesy=cmsy5
\font\eightit=cmti8
\font\eightsl=cmsl8
\font\eighttt=cmtt8
\font\tenfrak=eufm10
\font\sevenfrak=eufm7
\font\fivefrak=eufm5
\font\tenbb=msbm10
\font\sevenbb=msbm7
\font\fivebb=msbm5
\font\tensmc=cmcsc10


\newfam\bbfam
\textfont\bbfam=\tenbb
\scriptfont\bbfam=\sevenbb
\scriptscriptfont\bbfam=\fivebb

\newfam\frakfam
\textfont\frakfam=\tenfrak
\scriptfont\frakfam=\sevenfrak
\scriptscriptfont\frakfam=\fivefrak


\def\eightpoint{%
\textfont0=\eightrm   \scriptfont0=\sixrm
\scriptscriptfont0=\fiverm  \def\rm{\fam0\eightrm}%
\textfont1=\eighti   \scriptfont1=\sixi
\scriptscriptfont1=\fivei  \def\oldstyle{\fam1\eighti}%
\textfont2=\eightsy   \scriptfont2=\sixsy
\scriptscriptfont2=\fivesy
\textfont\itfam=\eightit  \def\it{\fam\itfam\eightit}%
\textfont\slfam=\eightsl  \def\sl{\fam\slfam\eightsl}%
\textfont\ttfam=\eighttt  \def\tt{\fam\ttfam\eighttt}%
\textfont\bffam=\eightbf   \scriptfont\bffam=\sixbf
\scriptscriptfont\bffam=\fivebf  \def\bf{\fam\bffam\eightbf}%
\abovedisplayskip=9pt plus 2pt minus 6pt
\belowdisplayskip=\abovedisplayskip
\abovedisplayshortskip=0pt plus 2pt
\belowdisplayshortskip=5pt plus2pt minus 3pt
\smallskipamount=2pt plus 1pt minus 1pt
\medskipamount=4pt plus 2pt minus 2pt
\bigskipamount=9pt plus4pt minus 4pt
\setbox\strutbox=\hbox{\vrule height 7pt depth 2pt width 0pt}%
\normalbaselineskip=9pt \normalbaselines
\rm}


\def\pagewidth#1{\hsize= #1}
\def\pageheight#1{\vsize= #1}
\def\hcorrection#1{\advance\hoffset by #1}
\def\vcorrection#1{\advance\voffset by #1}

\newcount\notenumber  \notenumber=1              
\newif\iftitlepage   \titlepagetrue              
\newtoks\titlepagefoot     \titlepagefoot={\hfil}
\newtoks\otherpagesfoot    \otherpagesfoot={\hfil\tenrm\folio\hfil}
\footline={\iftitlepage\the\titlepagefoot\global\titlepagefalse
           \else\the\otherpagesfoot\fi}

\def\abstract#1{{\parindent=30pt\narrower\noindent\eightpoint\openup
2pt #1\par}}
\def\smc{\tensmc}


\def\note#1{\unskip\footnote{$^{\the\notenumber}$}
{\eightpoint\openup 1pt
#1}\global\advance\notenumber by 1}

\def\frac#1#2{{#1\over#2}}

\def\tfrac#1#2{{\textstyle{#1\over#2}}}
\def\({\left(}
\def\){\right)}
\def\<{\langle}
\def\>{\rangle}
\def\2pd#1#2#3{\frac{\partial^2#1}{\partial#2\partial#3}}

\def\sqr#1#2{{\vcenter{\vbox{\hrule height.#2pt
        \hbox{\vrule width.#2pt height#1pt \kern#1pt
           \vrule width.#2pt}
        \hrule height.#2pt}}}}
\def\square{\mathchoice\sqr64\sqr64\sqr{4.2}3\sqr33}
\def\ni{\noindent}
\def\ref #1{$^{[#1]}$}
\def\slash{\!\!\!\!/}
\def\lqq{\lq\lq}
\def\rqq{\rq\rq}

\def\d{\delta}
\def\phys{{\hbox{\sevenrm phys}}}
\def\L{{\cal L}}
\def\psiphys{\psi_\phys}


\pageheight{24cm}
\pagewidth{15.5cm}
\hcorrection{-2.5mm}
\magnification \magstep1
\baselineskip=16pt
\parskip=5pt plus 1pt minus 1pt
%
%
\rightline {MZ-TH/93-05}
\rightline {DIAS-STP-93-05}
\vskip 40pt
\centerline{\bigbold ON THE PHYSICAL PROPAGATORS OF QED}
\vskip 30pt
\centerline{\smc Martin Lavelle{\hbox {$^*$}}{\note{e-mail:
lavelle@vipmza.physik.uni-mainz.de}}
and  David McMullan{\hbox {$^{\dag}$}}{\note{e-mail:
mcmullan@stp.dias.ie}}}
\vskip 15pt
{\baselineskip 12pt \centerline{\null$^*$Institut f\"ur Physik}
\centerline{Johannes Gutenberg-Universit\"at}
\centerline{Staudingerweg 7, Postfach 3980}
\centerline{W-6500 Mainz, F.R.\thinspace Germany}
\vskip 12pt
\centerline{\null$^{\dag}$Dublin Institute for Advanced Studies}
\centerline{School of Theoretical Physics}
\centerline{10 Burlington Road}
\centerline{Dublin 4}
\centerline{Ireland}
}
\vskip 7truemm
\vskip 40pt
{\parindent=0.58in\narrower\baselineskip=13pt\ni{\bf Abstract}\quad
The true variables in QED are the transverse photon components and
Dirac's physical electron, constructed out of the fermionic field
and the longitudinal components of the photon. We calculate the
propagators in terms of these variables to one loop and demonstrate
their gauge invariance. The physical electron propagator is shown
not to suffer from infrared divergences in any gauge. In general, all
physical Green's functions are gauge invariant and infrared-finite.
\bigskip\bigskip
\par}

\vfill\eject
\ni
The usual covariant formulation of QED exhibits a gauge dependence which
implies that the fermion in the Lagrangian and the four components of
$A_\mu$ cannot be physical quantities. In this letter we discuss
the physical fields and calculate their propagators in lowest order
perturbation theory.

In a general Lorentz gauge the QED Lagrangian is
$$
\L=-\tfrac14 F_{\mu\nu}F^{\mu\nu} - \tfrac1{2\xi} (\partial_\mu A^\mu)^2
+\bar\psi(i D\slash-m)\psi + i\bar c\, \square\,   c\,,
\eqno (1)
$$
where $D_\mu=\partial_\mu+ i g A_\mu$ and the ghosts are Hermitian.
This Lagrangian clearly does not only depend on the physical degrees
of freedom, thus some mechanism is needed to isolate the true
physical states. The Lagrangian (1) has, however, the
following BRST invariance
$$
\eqalign{\d A_\mu & =  \partial_\mu c\,,\cr
\d c & =  0\,,\cr
\d\bar c& =  -\tfrac{i}\xi \partial_\mu A^\mu\,,\cr
\d\psi& =-i g c\psi\,,\cr
\d\bar\psi&=-i g \bar\psi c\,,
}\eqno (2)
$$
which is generated by the conserved, nilpotent BRST charge
$$
Q =\int d^3x \bigl((\partial_i F^{0i}+gJ_0)c - \frac1\xi \partial_\mu
A^\mu \dot c\bigr)\,.
\eqno (3)
$$
Those states $|\psi\>$ for which $Q|\psi\> =0$, which are not of the
zero norm type $|\psi\>=Q|\chi\>$ (recall $Q^2=0$), can be
identified\ref{1} with the physical states of QED.  Some of the
complications encountered in excluding
these zero norm states can be avoided by the use of a further
symmetry of QED which we have recently observed and which may be
directly employed to isolate the physical states\ref{2}.
The result of this analysis is that the physical degrees
of freedom are just the two transverse photon components and
Dirac's physical electrons\ref{3}
$$
\psiphys(x)=    \psi(x)
\exp(ig\frac{\partial_i A_i(x)}{\nabla^2})\qquad
\hbox{and} \qquad
\bar \psiphys(x)=\bar\psi(x)\exp(-ig\frac{\partial_i A_i(x)}{\nabla^2})\,.
\eqno (4)
$$
The non-local
exponential function creates the Coulomb change in the electric field
around the electron\ref{3}. Since $\psiphys$ represents an electron with
spin, mass {\it and} charge, we will call this the electron and
$\psi$ the fermion in this letter. These fields are
physical and hence their Greens functions must also be gauge invariant.
The purpose
of this letter is to start to develop a perturbation theory
for these physical variables: we will do this by calculating their
propagators to one loop. These will be seen to yield a
gauge invariant, infrared-finite description of QED.
\smallskip
The physical photon, $A_\mu^\phys$, is given by
$$
A_\mu^\phys(k)=P_{\mu\nu}(k)A^\nu(k)\,,\eqno (5)
$$
where
$$
P_{\mu\nu}(k)=
\left[ g_{\mu\nu} + \frac{k_\mu k_\nu + k^2 \eta_\mu \eta_\nu - (k_\mu
\eta_\nu +
k_\nu \eta_\mu) k\cdot\eta}{k^2-(k\cdot\eta)^2} \right]\,,
\eqno(6)
$$
and $\eta$ is the temporal vector, $(1,0,0,0)$. This projector
satisfies $P_{0\nu}=k^\mu P_{\mu\nu}=0$, and so $A^\phys_\mu$
has just two components, the transverse components of the photon.
The physical photon propagator is then
$$
D_{\mu\nu}^\phys(k)= \int\!\frac{d^4k}{(2\pi)^4}\<T\bigl(A_\mu^\phys(x)
A_\nu^\phys(y)\bigr)\> e^{ik\cdot(x-y)}\,.
\eqno(7)
$$
It must be gauge invariant.
We can easily see that this is the case for the free propagator which
is

$$
\openup1\jot
\eqalign{
D^{(0)\hbox{\phys}}_{\mu\nu}(k)&=P_{\mu\lambda}(k)D^{\lambda\sigma\,(0)}(k)
P_{\sigma\nu}(k)\cr
&=-\frac 1{k^2} \left[ g_{\mu\nu} +
\frac{k_\mu k_\nu + k^2 \eta_\mu \eta_\nu - (k_\mu \eta_\nu +
k_\nu \eta_\mu) k\cdot\eta}{k^2-(k\cdot\eta)^2} \right]
\,.
}\eqno (8)
$$
where the bracketed superscript refers to the power of the coupling
and $D^{\lambda\sigma\,(0)}(k)$ is the full free propagator from~(1).
This gauge invariant result is just
the free propagator in radiation gauge\ref{4}.

To one loop,
Fig.\thinspace 1, the propagator receives the following
contribution
$$\openup1\jot
\eqalign{
D^{(2)\hbox{\phys}}_{\mu\nu}(k)&= P_{\mu\rho}(k)D^{\rho\lambda\,(0)}(k)
\Pi_{\lambda\sigma}^{(2)}(k)D^{\sigma\tau\,(0)}(k) P_{\tau\nu}(k)\cr
&=\frac{1}{k^2}
\left[ g_{\mu\nu} + \frac{k_\mu k_\nu + k^2 \eta_\mu \eta_\nu - (k_\mu
\eta_\nu +
k_\nu \eta_\mu) k\cdot\eta}{k^2-(k\cdot\eta)^2} \right]\Pi(k)
\,,}
\eqno(9)
$$
where $\Pi^{(2)}_{\lambda\sigma}(k)$ is the one loop photon
polarisation tensor and $\Pi(k)$ is the standard scalar polarisation
to one loop. This result is gauge invariant and is again just what one
would obtain
in radiation gauge. We stress that, contrary to some suggestions in the
literature, it is {\it not} necessary to go to
radiation gauge to obtain this result since this projection on
the propagator is
gauge invariant.
Indeed from the Ward identity we know that the photon polarisation must
have the tensor structure \hbox{$g_{\mu\nu}-\tfrac{k_\mu
k_\nu}{k^2}$} to all
orders in the coupling. However, from the transversality to the
momentum~$k$
of both the polarisation and of the projector $P$ on the physical photons
the gauge dependence in the photon propagators vanishes
in the physical propagator. Any gauge dependence in the scalar polarisation
would now directly enter the physical propagator as a multiplicative
factor. We thus see that the photon polarisation must be
independent of the gauge parameter to all orders in the coupling\ref{5}.
This may
be contrasted with the full propagator, where the gauge dependence resides
in the propagator for longitudinal photons.
\bigskip
The electron propagator is
$$
S^\phys(k)=\int\!\frac{d^4k}{(2\pi)^4}\<T\bigl( \psiphys(x)
\bar\psiphys(y)\bigr)\>e^{ik\cdot(x-y)}\,,
\eqno(10)
$$
which implies the free propagator
$$
S^{\phys\,(0)}(k)=\frac i{p\,\slash + m}\,.
\eqno(11)
$$
This is, of course, the usual, gauge invariant, free fermion
propagator ---
reflecting the fact that the fermion is identical to the true
electron at $\hbox{\cal O}
(g^0)$, or, to put it another way, that the fermion is  BRST invariant in
lowest order in the coupling.

To one loop (see Fig.\thinspace2) there are four separate
contributions to the electron propagator: the usual self-energy
contribution, i.e., the
exponentials are treated to lowest order in the coupling,
cross terms where one exponential factor is expanded to order
$g$ and a term where both exponentials are so expanded. Note that
if either of the exponentials is expanded to second order in the
coupling we
obtain a tadpole that vanishes in dimensional regularisation. The
terms coming from the exponential factors describe interactions with
longitudinal photons.

As the reader
can see from Fig.\thinspace2, we now have to directly calculate the
propagator itself; the final three diagrams do not let
themselves be interpreted as self energy terms. There are two
ways of calculating this propagator. In the Lorentz class all four diagrams
contribute. In Coulomb gauge the exponentials are unity and only the self
energy diagram survives. (Equivalently one could calculate in radiation
gauge where the exponentials are also trivial; there an extra diagram from
the four fermion interaction appears, which in the not completely physical
Coulomb gauge is duplicated by the contribution to the
self energy from the exchange of temporal
photons.) Taking for simplicity massless QED it can
be seen by direct calculation of the diagrams in Fig.\thinspace2, in the
full Lorentz class, that the one loop electron
propagator becomes
$$
S^{\phys\,(2)}(k)=-\frac{g^2}{k\,\slash}\int
\frac{d^Dp}{(2\pi)^D}\gamma_\mu\frac{p\,\slash-k\,\slash}{(p-k)^2}
\gamma_\nu
\frac1{p^2}\left[
-g^{\mu\nu}+\frac{p^\mu p^\nu-\eta{\cdot}
p(p^\mu\eta^\nu+p^\nu\eta^\mu)}{p^2-(\eta{\cdot} p)^2}
\right]\frac1{k\,\slash}\,.\eqno(12)
$$
All gauge parameter dependent terms from the various diagrams have,
as expected, cancelled.
In (12) we recognise
$$
\frac1{p^2}\left[
-g^{\mu\nu}+\frac{p^\mu p^\nu-\eta{\cdot }
p(p^\mu\eta^\nu+p^\nu\eta^\mu)}{p^2-(\eta{\cdot} p)^2}
\right]=D_{\mu\nu}^{\hbox{\sevenrm Coul}\,(0)}(p)\,,
\eqno(13)
$$
the free propagator in the Coulomb gauge. This further demonstrates
the gauge invariance of this result, since in Coulomb gauge the
electron and fermion propagators are equal. The fermion propagator
to one loop in Coulomb gauge may be found in Ref.\thinspace 6.

This analysis may be straightforwardly extended to higher Green's
functions. For the vertices this means explicitly that the physical
electron transverse photon vertex,
$\< \psiphys(x) \bar\psiphys(y) A^\phys_\mu(z)\>$, is
invariant and that unphysical vertices, like that with the temporal photon
component $\< \psiphys(x) \bar\psiphys(y) A_0(z)\>$, are not. From the
gauge invariance of the physical vertex it must be equivalent to
the one calculated in Coulomb gauge, where electrons may be
replaced by fermions: i.e., the physical vertex calculated
in any gauge will yield the Coulomb gauge result\ref{6}
for $\< \psi(x) \bar\psi(y) A^\phys_\mu(z)\>$.

We have in other words the following equivalent recipes for
physical Green's functions in QED,
i.e., those defined in terms of $\psiphys$, $\bar\psiphys$ and
the transverse photons, $A^\phys_\mu$. One may either calculate
them in any gauge, say a Lorentz gauge, directly, or one may work in
Coulomb or radiation gauge where one may use the usual fermions,
$\psi$. The end results will be identical. (Note that in Coulomb
gauge one must still project out the external photons upon the physical
sector, with that gauge
condition this is simply a matter of dropping external temporal photons.)

One of the attractions of Coulomb gauge is its infrared finiteness, and
we see that all gauges will now share this important property
if physical fields are used.
\bigskip
We have seen above that it was possible to construct a sensible
perturbation theory for the physical fields of QED despite their
non-local nature. The extension of this approach to vertex corrections
and higher loops should not prove significantly more difficult than
the usual version of QED even if covariant gauges are employed.
Indeed this description, although at
first sight appearing more cumbersome than the standard covariant
formulation, has many attractive features: the physical Green's functions
are explicitly gauge invariant and the S-matrix is already
infrared-finite.

Acting on the vacuum the operator corresponding to Dirac's electron
(4) will create a state outside of the Fock space. This is simply a
consequence of the exponential factor. Coherent states provide a
natural arena to discuss this state. Thus we see that basing QED on
these physical fields produces an infrared-finite theory, at the
expense of extending the traditional Fock space to include coherent
states upon which these operators are defined. This should be
contrasted with the traditional approach to QED, where the infrared
divergences of the Green's functions motivate the introduction of
coherent states\ref{7}. A fuller account of the connection between
the coherent states  of the physical fields and those generally used
will be provided elsewhere\ref{8}.

Although, as we have recently shown\ref{9}, the Gribov ambiguity
prevents the construction of a physical quark, it is locally, i.e.,
perturbatively, possible. (This pseudophysical quark will not, however,
reduce in Coulomb gauge to the Lagrangian
fermion.) Such perturbative quarks are \lq
seen\rq\ in deep inelastic scattering. The non abelian extension of
this letter would thus provide a gauge invariant description of the
propagation and interaction of perturbatively physical quarks in
deep inelastic scattering.
\bigskip
\bigskip
\ni{\bf Acknowledgements} MJL thanks the Dublin Institute for Advanced
Studies for their warm hospitality and the Graduierten Kolleg of Mainz
University for support.
\vfill
\eject
\ni{\bf Figure Captions}
\itemitem{\bf Fig.\thinspace1}{The one loop contribution to the photon
propagator.}
\itemitem{\bf Fig.\thinspace2}{One loop contributions to the electron
propagator. The dashed lines represent contributions from the longitudinal
components of $A_i$ coming from the expansion of the exponential factors.}
\bigskip
\ni{\bf References}
\item{1)}{N. Nakanishi and I. Ojima, \lqq Covariant Operator Formalism of
Gauge Theories and Quantum Gravity\rqq, (World Scientific,
Singapore, 1990).}
\item{2)}{M. Lavelle and D. McMullan, \lqq A New Symmetry for QED\rqq,
Mainz/Dublin preprint, MZ-TH/93-02, DIAS-STP-93-03. Submitted to Physical
Review Letters.}
\item{3)}{P.A.M. Dirac, \lqq Principles of Quantum Mechanics\rqq, (OUP,
Oxford, 1958), page 302.}
\item{4)}{J.B. Bjorken and S.D. Drell, \lqq Relativistic Quantum
Fields\rqq, (McGraw-Hill, New York 1965), Sect.\thinspace 14.6;
M.~Lavelle and D.~McMullan, \lqq The Radiation Class: A Set of
Temporal Gauges\rqq, to appear in Z. Phys. C.}
\item{5)}{V.B. Berestetskii, E.M. Lifshitz and L.P. Pitaevskii, \lqq
Quantum Electrodynamics\rqq, (Pergamon Press, Oxford, 1982),
Sect.\thinspace103.}
\item{6)}{G.S. Adkins, Phys. Rev. D27 (1983) 1814.}
\item{7)}{T.W.B. Kibble, in \lqq Mathematical Methods in Theoretical
Physics\rqq, ed.s K.T. Mahanthappa and W.E. Brittin (Gordon and
Breach, New York, 1969).}
\item{8)}{M. Lavelle and D. McMullan, \lqq Physical States and
Greens Functions in Gauge Theories: I QED\rqq, in preparation.}
\item{9)}{M. Lavelle and D. McMullan, \lqq On Quark Confinement\rqq,
Mainz/Dublin preprint, MZ-TH/93-02, DIAS-STP-93-03. Submitted to
Physical Review Letters.}

\bye